# Wavefront manipulation based of the excitation of bound states in dielectric photonic crystals and metasturctures

Anna C. Tasolamprou[1], Maria Kafesaki[1,2], Thomas Koschny[3] Costas M. Soukoulis[1,3]

1. Institute of Electronic Structure and Laser, FORTH, 70013 Heraklion, Crete, Greece
2. Department of Materials Science and Technology, University of Crete, 70013, Heraklion, Crete, Greece
3. Ames Laboratory and Department of Physics and Astronomy, Iowa State University, Ames, Iowa 50011, United States
*corresponding author, E-mail: atasolam@iesl.forth.gr

## Abstract

We present the study of bound surface modes sustained at the termination of truncated bulk dielectric photonic crystals and isolated metasurfaces of dielectric meta-atoms. We discuss the origins of bound modes in the two systems and their relation. For both systems, we theoretically study and experimentally demonstrate wavefront manipulation, in particular directional emission, frequency splitting and beam collimation achieved by coupling of the bound states to radiation modes through leaky wave radiation mechanism using properly designed scattering gratings.

## 1. Introduction

Surface modes are non-radiating states that are found bound to the interface of different media. Traditionally bound states have been investigated at the interface of semi-infinite dielectric and metallic interfaces at the optical frequencies, the known, so-called surface plasmon polaritons (SPP). SPPs offered a very good platform for highly confining propagating waves and may enable the minimization of electromagnetic components. However, at the same time, dielectric/metallic SPPs suffer from high dissipation losses which pose various application restrictions. However, it was also known that bound states can also be supported in dielectric media. These bound states stem from the excitation of coupled Mie resonances in the dielectric system and lead to light confinement (Mie modes are generally more spatially extended than SPPs). The fact that dielectric media suffer no dissipation losses has brought the non-radiating bound states at the center of attention of composite, all-dielectric resonant structures and metamaterials [1]. In this work we present two type of structures that sustain surface modes, one consists of a metasurface of dielectric meta-atoms [2] and the other consists of truncated photonic crystals [3]. In the first case we demonstrate experimentally that beam collimation can be sustained at large distance via a system of cascading dielectric metasurfaces and in the other case we present the controlled directional emission engineered at the exit of photonic crystal line defect waveguide. For both cases we present the corresponding experimental demonstration which is conducted in the microwave regime and provides a proof of concept verification our analysis. The dielectric structures are be scaled down for the optical/NIR regime [4].

Additionally we will present a theoretical framework that analyses the nature and the origins of the bound surface states in the termination of truncated bulk photonic crystals and their relation to bound states sustained in isolated dielectric metasurfaces.

## 2. Discussion

The first structure under consideration consists of a system of dielectric metasurface bilayers; the first layer of each bilayer supports surface states and the rear, the grading layer couples the surface states to radiation modes. The first layer of each bilayer consists of 35 alumina circular rods with a lattice constant of $\alpha = 11$ mm, while the second layer of square alumina rods with lattice constant $b = 2\alpha$. The experimental set-up consists of an HP E8364B network analyser, a horn antenna as the transmitter and a dipole antenna as the receiver. The horn antenna transmits a Gaussian beam with E polarization (the electric field parallel to the dielectric rods) and the dipole antenna scans the 2D experimental table, measuring the intensity of the local field close to the structure and the propagation, diffraction and interference of the outgoing waves. In Figure 1(a) and Figure 1(b) we present strength distribution in a multi-bilayer metasurface system and it has been experimentally verified. We also observe Fabry-Perot type standing wave oscillations that occur in the space that lies between the cascading metasurfaces. Nevertheless, beam collimation at large distances is observed. Figure 1(c) plots the normalized field distribution of the two structures at a propagation distance equal to 2.6 m ($100\lambda$). It is obvious that four cascading metasurfaces are capable of cancelling the divergence. The second type of the structure consists of a photonic crystal terminate by a dielectric layer of the same periodicity and different size than the bulk photonic crystal. The dispersion diagram of the surface state in the bulk termination, shown in red, stands within the photonic crystal bandgap and below the light line which indicates that the mode is dark, bound between in the surface layer the bulk photonic crystal and the air as it is presented in Figure 2. It is in fact almost identical to that of the isolated metasurface of the dielectric of rods. This indicates that the origins of the surface states in the both system are similar. A line defect waveguide formed in the bulk photonic crystal feeds the dark mode that propagates to the exit sides of the structure. A grating layer with double periodicity placed after the surface layer undertakes the coupling of the surface modes into forward radiation. Forcing a small asymmetry in the grating layer leads to the frequency selective modification of the angle of the forward propagating waves. The angle of the emission obeys the surface mode dispersion and grating equation that describes the leaky wave radiation mechanism. In Figure 3 we present the experimentally measurement and numerical study of the near and intermediate field exciting the line defect waveguide. For the experimental study we use the same set-up as that described in the discussion of Figure 1. Oblique directionality at different frequencies is observed enabling the frequency splitting operation. Similar operation can be achieved in silicon based inverse photonic crystal for operation in the near infrared and optical regime. The corresponding structure and operation is thoroughly discussed in Ref. [4].

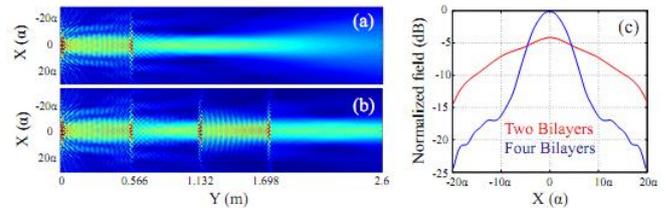

Figure 1: Simulation of the 2D field strength at the exit of a system of bilayer dielectric metasurfaces at the resonant frequency: case of (a) two metasurfaces and (b) four metasurfaces structure. (c) Normalized x cross section field distribution for the two cases at propagation distance $100\lambda$.

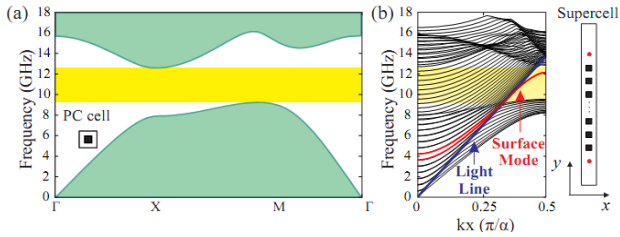

Figure 2: Dispersion diagram in a bulk, infinite dielectric photonic crystal of dielectric rods. The dispersion curves are calculated via the plane wave expansion method for the infinite periodic structure. (b) Dispersion curves for the finite PC along with the surface layer. They are calculated by applying the plane wave expansion in the supercell shown above. The surface mode dispersion is plotted in red; it lies within the badgap and below the light line (blue curve), and occurs at 10.1−12.2 GHz.

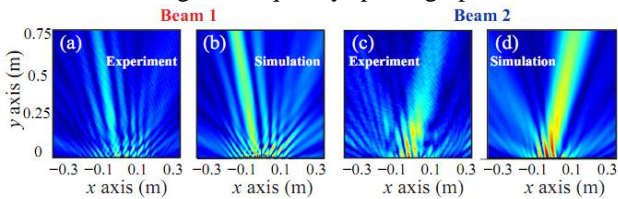

Figure 3: Experimental (a) and simulated (b) 2D plot of the strength of the outgoing near and intermediate field at the exit of a photonic crystal terminated by a surface layer and an asymmetric grating layer at frequency $f^{exp1}$ = 11.70 GHz and $f^{sim1}$ = 11.68 GHz. Axis $x$ is parallel to the grating layer and axis $y$ perpendicular to the grating layer. Experimental (c) and simulated (d) 2D plot of the intensity of the outgoing near and intermediate field at frequency $f^{exp2}$ = 10.20 GHz and $f^{sim2}$ = 10.18 GHz.

## 3. Conclusions

We studied theoretically and demonstrated experimentally the realization of wavefront manipulation functions based on the excitation and handling of dark bound surfaces states in dielectric resonant media such as photonic crystals and periodic metasurfaces made of dielectric meta-atoms. In particular, we have demonstrated the sustaining of the beam collimation is possible in a system of cascading dielectric metasurfaces. Additionally we showed that frequency selective, directional emission from a line defect photonic crystal waveguide can be attained if the bulk photonic crystal is properly terminated. The characteristics of the bound states are similar for the two systems indicative of their common origins. Experiments have been conducted in the microwave regime and the structures can be directly scaled in the near infrared and optical frequencies. Due to the absence of ohmic losses, the non-radiating bound states provide a promising platform for the manipulation of light attract the growing attention of the nanophotonic community.

### Acknowledgements

This work was supported by the European Research Council under ERC Advanced Grant no. 320081 (project PHOTOMETA) and the European Union's Horizon 2020 Future Emerging Technologies call (FETOPEN-RIA) under grant agreement no. 736876 (project VISORSURF). Work at Ames Laboratory was supported by the US Department of Energy (Basic Energy Science, Division of Materials Sciences and Engineering) under Contract No. DE-AC02-07CH11358.